\documentclass[twocolumn]{aastex63}
\usepackage[]{units}
\usepackage[update,prepend]{epstopdf}
\usepackage{graphicx}
\usepackage{amssymb}
\usepackage{amsmath}
\usepackage{threeparttable}
\usepackage{hyperref}
\usepackage{color}

\usepackage{mathtools}

\usepackage{makecell}

\newcommand{\kms} {\,km\,s$^{-1}$}

\mathchardef\mhyphen="2D
\shorttitle{Very wide companions around hot jupiter hosts and contact binaries}
\shortauthors{Hwang et al.}

\graphicspath {{figs/} }

\begin{document}

%\title{Enhanced wide companion fraction around hot jupiter hosts from Gaia DR2}
\title{Very wide companion fraction from Gaia DR2: a weak or no enhancement for hot jupiter hosts, and a strong enhancement for contact binaries}

\author[0000-0003-4250-4437]{Hsiang-Chih Hwang}
\affiliation{Department of Physics \& Astronomy, Johns Hopkins University, Baltimore, MD 21218, USA}

\author[0000-0002-7993-4214]{Jacob H. Hamer}
\affiliation{Department of Physics \& Astronomy, Johns Hopkins University, Baltimore, MD 21218, USA}

\author[0000-0001-6100-6869]{Nadia L. Zakamska}
\affiliation{Department of Physics \& Astronomy, Johns Hopkins University, Baltimore, MD 21218, USA}

\author[0000-0001-5761-6779]{Kevin C. Schlaufman}
\affiliation{Department of Physics \& Astronomy, Johns Hopkins University, Baltimore, MD 21218, USA}

\begin{abstract}

There is an ongoing debate on whether hot jupiter hosts are more likely to be found in wide binaries with separations of $\gtrsim 100$\,AU. In this paper, we search for comoving, very wide companions with separations of $10^3-10^4$\,AU for hot jupiter hosts and main-sequence contact binaries in Gaia DR2, and compare the very wide companion fractions with their object-by-object-matched field star samples. We find that $11.9\pm 2.5$\% of hot jupiter hosts and $14.1\pm 1.0$\% of contact binaries have companions at separations of $10^3-10^4$\,AU. While the very wide companion fraction of hot jupiter hosts is a factor of $1.9\pm0.5$ larger than their matched field star sample, it is consistent, within $\sim1\sigma$, with that of matched field stars if the matching is only with field stars without close companions (within $\sim50$\,AU) as is the case for hot jupiter hosts. The very wide companion fraction of contact binaries is a factor of $3.1\pm0.5$ larger than their matched field star sample, suggesting that the formation and evolution of contact binaries are either tied to or correlated with the presence of wide companions. In contrast, the weak enhancement of very wide companion fraction for hot jupiter hosts implies that the formation of hot jupiters is not as sensitive to those environment properties. Our results also hint that the occurrence rates of dual hot jupiter hosts and dual contact binaries may be higher than the expected values from random pairing of field stars, which may be due to their underlying metallicity and age dependence.
\end{abstract}

%\red{()NLZ: Do we want more? Another sentence in the beginning or in the end on why that's relevant? What is the mechanism that would produce Hot jupiters and how does the excess of companions at 1e3-1e4 possibly contributing? Is it the same or different from the contact binaries? If different, is the factor of 3 a coincidence?..)

\section{Introduction}

%\blue{Jacob: intro on Hot jupiters? (a) Why hot jupiters are interesting? (b) Formation (or orbital migration) of hot jupiters, for example some people suggest that the Kozai mechanism can explain their formation (Fabrycky \& Tremaine 2007), and the current consensus. Why looking into companion fraction is important. (c) Debate on the fraction of wide companions around hot jupiters (Ngo+16, Moe+19), and here we report an enhanced fraction.}

The discovery of the first exoplanet orbiting a sun-like star, 51 Pegasi b, presents a significant challenge to planet formation theories based on our Solar System (e.g. \citealt{Mayor1995, Rasio1996}). While the giant planets of the Solar System may have somewhat migrated \citep{Morbidelli2012}, they likely formed in the cooler outer regions of the protoplanetary disc where the icy material facilitated rapid core growth to accrete massive gaseous atmospheres before the disk dissipated. In contrast, 51 Pegasi b is the prototypical `hot jupiter', with a mass roughly half that of jupiter but with an orbital separation  from the host star of about 7 times smaller than that of Mercury. 25 years after the discovery of 51 Pegasi b, the formation of hot jupiters remains an open question.
%with a mass roughly half that of jupiter but an orbital period of only 4.23 days

%In our solar system, the formation of the giant planets exterior to the rocky planets could be explained by icy material in the cooler outer regions of the protoplanetary disc facitiating more rapid planetesimal growth.

There are three main hot jupiter formation theories: in situ formation, disk migration, and high-eccentricity tidal migration \citep{Dawson2018}. In the in situ formation scenario, hot jupiters form at their current locations. This scenario was believed to be challenging because both gravitational instability and core accretion are difficult to operate at hot jupiters' close-in locations \citep{Rafikov2005, Rafikov2006}, but recent study suggests that in situ formation is still possible under certain conditions \citep{Bodenheimer2000,Batygin2016, Boley2016, Lee2016a}. In the disk migration scenario, hot jupiters form at larger separations and migrate to their current location under the torque from the protoplanetary disk \citep{Goldreich1980, Lin1986, Lin1996, Nelson2000, Ida2008}. 

%In this scenario, migration is efficient and rapid \citep{Nelson2000} and \red{difficult to halt at the typical orbital distances of hot jupiters before they fall into the star} \citep{Vidotto2009}.

In the high eccentricity migration scenario, the hot jupiter forms at a large separation from the host star, is driven into a high-eccentricity orbit, and undergoes tidal circularization which leaves the planet in its small, circular orbit. The eccentricity excitation may be caused by planet-planet scattering \citep{Rasio1996a,Weidenschilling1996,Ford2006,Chatterjee2008, Juric2008}, or Kozai-Lidov interactions \citep{Kozai1962, Lidov1962} with an other planet \citep{Naoz2011} and/or with a stellar companion \citep{Wu2003,Fabrycky2007, Naoz2012}. 
%\red{(NLZ: This is a little weird. You mention three scenarios, but then only talk about one. Add one sentence each / one reference each for the other ones?.. )}

%The focus of this work will be on the HEM channel of formation. 

%While no single formation channel can explain all properties of hot jupiters and multiple formation channels may be responsible, the high-eccentricity formation scenario \citep{Dawson2018}.

%Evidence suggests that the high-eccentricity tidal migration scenario must play some rule in the formation of hot jupiters. For example, the existence of moderately eccentric ($e\gtrsim0.2$) hot jupiters cannot be easily explained by the in situ formation and the disk migration, but is expected from the high-eccentricity tidal migration where the eccentric hot jupiters are on the way to circularization \citep[e.g.][]{Matsumura2010,Pont2011,Husnoo2012}. 

The occurrence of stellar companions around hot jupiter hosts provides a constraint on the formation of hot jupiters. If Kozai-Lidov interactions with stellar companions represent a significant channel for hot jupiter formation, then stellar companions should be common around hot jupiter hosts. At an earlier evolutionary stage, the presence of stellar companions can affect the environment of the protoplanetary disk and the planet formation \citep{Kraus2012}.
%\red{At an earlier evolutionary stage, the stellar companions can affect the environment of the planet formation site by inducing the streaming instability} \citep{Youdin2005, Johansen2007}.

%then hot jupiter hosts should have a higher companion fraction than similar stars without hot jupiters. The measurement of the fraction of stellar companions may also give us an insight into whether the formation of hot jupiters is tied to binary formation. 

%Not only do stellar companions may help the proto-hot jupiters to migrate, stellar companions can also change the environment of the planet formation site dramatically.

There has been extensive work on measuring the incidence of stellar companions to hosts of hot jupiters \citep{Knutson2014, Endl2014, Piskorz2015, Bryan2016, Evans2016, Belokurov2020}, but whether hot jupiter hosts have a higher wide companion fraction is still an ongoing debate. \citet{Ngo2016} conduct a direct imaging search for companions to hot jupiter systems and find that for companions with separations between 50 and 2000\,AU, hot jupiter hosts have a companion fraction 2.9 times higher than that of the field stars from \cite{Raghavan2010}, with a significance of 4.4\,$\sigma$. They argue that $>80$\% of these companions are not able to induce Kozai-Lidov oscillations because the oscillation timescale is too long, and therefore the enhanced companion occurrence may instead be linked to the formation environment of the gas giants.

% 4.4$\sigma$ higher than field stars. \red{(NLZ: I wouldn't phrase it in terms of sigma. You have an abstract in front of the reader where you have excess of a factor of 3. Let the reader compare your values with Ngo. "have a companion excess higher by ??\% than ... of by a factor of ??)} For companions between 1-50 AU, they found that hot jupiter hosts have a companion fraction 2.7$\sigma$ lower than the field. 

However, \cite{Moe2019b} point out that such enhanced companion fraction of hot jupiter hosts may be a consequence of several selection effects. First, they argue the field star sample from \cite{Raghavan2010} is not ideal for the hot jupiter hosts used in \cite{Ngo2016} because these two samples have slightly different mass and metallicity, and the field star sample is not complete.

%\blue{Second, \cite{Moe2019b} point out a selection effect that is frequently omitted in the literature. When a sample of hot jupiter hosts is selected, such selection already avoids close ($<50$\,AU) stellar companions because these close stellar companions suppress the planet formation. Indeed, observationally \cite{Ngo2016} find that only $4\%^{+4\%}_{-2\%}$ have companions within 50\,AU. If only close stellar companions suppress the planet formation and wider companions have no effects on it, then the enhanced wide companion fraction of hot jupiter hosts may be just a consequence of their reduced fraction of having short-separation companions. After accounting for these effects, \cite{Moe2019b} show that hot jupiter hosts do not have an enhanced wide companion fraction while short-period binaries still have an enhanced wide companions compared to the field stars. }

Second, observations have shown that fewer hot jupiter hosts have close stellar companions at separations $\lesssim50$\,AU compared to the field stars \citep{Wang2014,Wang2014a,Kraus2016,Matson2018,Ziegler2019}. This may be the result of the shorter life-time of protoplanetary disks in binaries compared to those around single stars \citep{Kraus2012, Kraus2016}. Specifically, hot jupiter hosts have a very small companion fraction within 50AU -- only $4\%^{+4\%}_{-2\%}$ \citep{Ngo2016}. In contrast, field stars with a similar mass and metallicity have a companion fraction within 50AU of $40\pm6$\% \citep{Moe2019b}. In the absence of any other physical mechanisms, the lower close companion fraction of hot jupiter hosts would result in a higher wide companion fraction compared to the field stars. After having accounted for this bias, \cite{Moe2019b} argue that hot jupiter hosts do not have an enhanced wide companion fraction compared to the field stars.

%\blue{If the stellar companions of the hot jupiter hosts and the field stars have the same separation distribution beyond} $50$\,AU, then the smaller companion fraction within $50$\,AU of hot jupiter hosts would make their wide companion fraction larger than the field stars.

%\red{For example, instead of "When a sample...": "Hot jupiter hosts have a very small companion fraction within 50AU -- only 4+4-2\% (Ngo et al. 2016). In contrast, field stars with a similar mass distribution have a companion fraction within 50AU of 40\% (ref). [Possible explanations for this difference? Observational biases? see my email. Physical mechanisms? Is there anything particularly interesting at the 50AU mark, physically?] Either way, it is not clear what is the proper comparison sample for hot jupiter hosts. Moe and Kratter (2019) suggest that the proper comparison is with field stars not having stellar companions within 50AU. Using this approach, they show that..."}

%They suggest that the mass dependence on the secondary is due to different formation scheme, where secondary masses $<4$\,M$_{\rm J}$ are formed from core accretion and secondary masses $>10$$_{\rm J}$ are formed from gravitational instability, coinciding with the mass gap found in \cite{Schlaufman2018}.

%In this paper, we use Gaia data to re-examine the companion fraction at separations $>1000$\,AU. 

In this paper, we examine the very wide companion fraction at separations of $10^3-10^4$\,AU around hot jupiter hosts and main-sequence contact binaries using Gaia data. We search for comoving and colocated companions to hot jupiter hosts, contact binaries, and to field stars down to Gaia's spatial resolution limit. By using Gaia, we are able to compare the companion fraction between different populations based on the same dataset, without dependence on external data and models. 

In this paper, we refer to companions with separations of $<50$\,AU as close companions, those of $50-2000$\,AU as wide companions, and those of $10^3-10^4$\,AU as very wide companions. This paper is structured as follows. Section~\ref{sec:selection} describes our sample selection and the search of comoving companions. Section~\ref{sec:result} presents our main results. We discuss the results in Section~\ref{sec:discussion} and conclude in Section~\ref{sec:conclusion}.

%\red{wide (We need to be a little careful. These are wide, but not as wide as ours... So in the paragraphs where you are talking about companions at 50-2000 AU, maybe replace the word ``wide'' with ``outer''? )}  

\section{Sample selection and method}
\label{sec:selection}

\subsection{Selection of hot jupiter hosts}

%\blue{Jacob will write this part. Especially, are they RV confirmed? Or they only have transit detections? It is possible that some hot jupiters are background eclipsing binaries? }

We use the sample of main sequence hot jupiter hosts whose selection is described in \citet{Hamer2019}. Briefly, hot jupiters were selected from the confirmed planets table of the NASA Exoplanet Archive using the fiducial definition from \citet{Wright2012}, planets having $P<10$ days and $M\sin{i}>0.1\ M_\mathrm{Jup}$. Most ($\sim90$\%) of these hot jupiters were discovered by transiting surveys and further confirmed by follow-up radial-velocity observations. Therefore, no contamination from false-positive eclipsing binaries is expected. This is important because an exoplanet sample from radial velocity surveys may be biased because exoplanet radial velocity surveys may exclude spectroscopic binaries \citep{Moe2019b}. In transiting surveys, the presence of a close companion may dilute the transit depths and potentially bias an exoplanet sample, but \cite{Moe2019b} show that hot jupiters are relatively immune to this effect because of their deep and frequent transits. The Gaia DR2 designations of these hot jupiter hosts were then obtained from SIMBAD. The sample was limited to hosts having good astrometry using the quality cuts described in Appendix 1 of \citet{Hamer2019}. Individual line-of-sight reddening values were calculated for each star by interpolating the three-dimensional reddening map from \citet{Capitanio2017} and integrating the interpolated grid along the line of sight to calculate a total $E$($B-V$) reddening. $E$($B-V$) was converted to Gaia reddening $E$(BP$-$RP) and extinction $A_{G}$, using the mean extinction coefficients from \citet{Casagrande2018a}. Evolved hot jupiter hosts were excluded by removing hosts which fall more than one magnitude above an empirical fit to the Pleiades in the $(\rm{BP-RP})$--$M_{G}$ plane. Among the 338 main-sequence hot jupiter hosts from \cite{Hamer2019}, we further limit the sample to parallaxes $>2.5$\,mas (distances $<400$\,pc) for better companion completeness. As explained in the later sections, we remove targets where no matched field stars are found or in a comoving group, ending up with 193 hot jupiter hosts.

%\red{NLZ: Somewhat confusing. You use passive tense and present tense both, so it's unclear what Hamer and Schlaufman did previously and what you are doing now. So I suggest refurbishing. Par 1: how did Hamer and Schlaufman select their hot- jupiter hosts? How many did they have? Passive tense. Par 2: what additional things do you do with their catalog in this paper? Present tense. How many objects you end up with. }

\subsection{Selection of main-sequence contact binaries}
\label{sec:gaia}

%\red{NLZ: There has been nothing about contact binaries so far. How are they relevant? You might add a few sentences in the introduction that you will not only compare hot jupiters to field stars, but also to contact binaries and why. You can say "Formation of contact binaries is puzzling in some of the same ways as the formation of hot jupiters, in that contact main-sequence binaries are closer together than their pre-main-sequence sizes, so their orbits must have considerably shrunk since their formation. Kozai-Lidov interactions between the binary and an outer stellar companion is often invoked to explain the short periods of contact binaries (Fabrycky and Tremaine). Therefore, in this paper we also compare the companion fraction between hot jupiter hosts, contact binaries and field stars." }

Short-period binaries are often compared to hot jupiters because historically both their formations are speculated to be due to the Kozai-Lidov interactions \citep[e.g.][]{Fabrycky2007}. Furthermore, a direct comparison of wide companion fractions between short-period binaries and hot Jupiters \citep{Moe2019b} provides a probe of the different formation processes across the mass gap of $4-9$\,M$_{\rm J}$ \citep{Schlaufman2018}. Therefore, in this paper we compare the companion fraction between main-sequence hot jupiter hosts, main-sequence contact binaries and field stars. 

%Also, their comparison provides insights in the different formation processes across the mass gap of $4-9$\,M$_{\rm J}$ \citep{Moe2019b}. Therefore, in this paper we compare the companion fraction between main-sequence hot jupiter hosts, main-sequence contact binaries and field stars. 

%\red{Rephrase and cite Kevin who wrote papers on the differences in formation across the gap. "Furthermore, a direct comparison of short-period binaries with hot Jupiters provides a probe of the different formation processes of binaries across a wide range of mass across the "brown dwarf desert" (refs). }

We use the sample of contact binaries whose selection is detailed in \cite{Hwang2020b}. They used the fractional variability from Gaia Data Release 2 (DR2; \citealt{Gaia2016,Gaia2018Brown}) to select high-amplitude ($>10$\%) variables, which are dominated by contact binaries on the main sequence. They selected main-sequence objects using $|\Delta \rm G| < 1.5$\,mag, where $\Delta \rm G$ is defined as the offset of absolute G magnitudes between the stars and Pleiades at the same BP$-$RP colors \citep{Hamer2019}. Other photometric and astrometric quality criteria employed in this work are the same as those used in \cite{Hwang2020b}. By comparing with the Kepler eclipsing binary catalog \citep{Kirk2016}, \cite{Hwang2020b} show that this method efficiently selects main-sequence contact binaries with orbital periods $<0.5$\,day. In this paper, we use a color cut of BP$-$RP$=$0.5-1.5\,mag to avoid blue pulsating stars and red M-dwarf flaring stars \citep{Gaia2019Eyer}.

With all selections discussed above, we further limit the main-sequence contact binary sample to parallaxes $> 2.5$\,mas (i.e. within 400\,pc), remove targets if we cannot identify a matched field star, and remove targets in comoving groups explained in the later sections. We end up with 1333 main-sequence contact binaries.

%All our targets have G $<16.2$\,mag, well above the Gaia DR2 detection limit of $\sim21$\,mag.

%We select main-sequence objects in the color range of BP$-$RP$=$0.5-1.5, where the high-amplitude variables are dominated by contact binaries. Our query for Gaia DR2 follows the one used in \cite{Gaia2018Babusiaux}. Specifically, the mean flux divided by its error is is larger than 50 for G-band and larger than 20 for BP and RP bands. In Gaia DR2, BP and RP fluxes are not treated with deblending, so we apply a cut on {\tt phot\_bp\_rp\_excess\_factor} to reduce the effect of crowded fields which makes the BP and RP bands unreliable \citep{Evans2018, Arenou2018}. {\tt visibility\_periods\_used}$>8$ is used to ensure that there are sufficient observations for deriving the astrometric solutions \citep{Lindegren2018}, and {\tt parallax\_over\_error}$>10$ is adopted to have well-measured parallaxes. We do not apply explicit cuts on relative proper motion uncertainties because that excludes objects having intrinsically low proper motions, which biases the kinematic results. Instead, we follow \cite{Gaia2018Babusiaux} to use the unit error introduced by \cite{Lindegren2018} to avoid spurious astrometric solutions.

\subsection{Control field star sample}

%To answer whether hot jupiter hosts and main-sequence contact binaries have higher \red{(higher than what)} wide-companion fractions, field star samples are needed for such comparison. 
For each target sample (hot jupiter host sample or main-sequence contact binary sample), we select a field star sample that matches object-by-object in several properties. Specifically, for each target star (hot jupiter host or a contact binary), we search a field star such that: (1) it is $>10$\,pc away from the target star (assuming the pair has the same parallax as the target star), and it is $<20$\,deg from the target star; (2) the BP$-$RP color difference is $<0.05$\,mag; (3) the parallax difference with the target star is $<10$\%; (4) the tangential velocity difference with the target star is $<10$\kms; (5) the field star satisfies the Gaia selections explained in Sec.~\ref{sec:gaia}. The tangential velocity used here has been corrected for the solar motion \citep{Schonrich2010} and the Galactic differential rotation using the Oort's constants reported by \cite{Bovy2017}. To improve statistics, for each hot jupiter host, we find four matched field stars, where the mutual separations of four matched field stars are all $>10$\,pc (assuming the same parallax as the target star). We exclude hot jupiter hosts from the sample if four matched field stars are not found. Because the contact binary sample is sufficiently large, only one field star is matched to one contact binary. Every control field star is matched to exactly one target.

By matching the properties described above, we ensure that the sample of the target stars (hot jupiter hosts or contact binaries) and the corresponding matched field star sample have similar distributions of masses, sky distributions, Galactic latitudes, and kinematics. Since the stellar age is strongly correlated with the kinematics \citep{Dehnen1998,Nordstrom2004,Reid2009,Sharma2014a}, we expect that the kinematic ages of the target sample and the matched field star sample are also similar. This is important because the occurrence of both hot jupiters and contact binaries has strong stellar age dependence \citep{Hamer2019,Hwang2020b}. Matching the Galactic latitudes is crucial because the contact binary fraction is much lower in the thick disk than in the thin disk \citep{Hwang2020b}.
%\blue{Since the stellar age is strongly correlated with the kinematics and metallicity is correlated with the Galactic height} \cite{Moe2019, Hwang2020a} \blue{(is Moe2019 relevant?)}, we expect that the distributions of metallicity and age are also similar between the target sample and the matched field star sample.

The hot jupiter occurrence rate is correlated with stellar metallicity \citep{Gonzalez1997, Santos2004, Fischer2005}. Therefore, ideally, we may also want to match metallicity for the control field stars. Although we match the masses, Galactic latitudes, and kinematics, which are all correlated with metallicity, there is no guarantee that the metallicity is exactly the same as the hot jupiter host sample. However, in this paper we focus on the very wide companions ($10^3-10^4$\,AU), and it is has been shown that the solar-type binary fraction beyond $>200$\,AU is independent of metallicity \citep{El-Badry2019a,Moe2019}. Therefore, metallicity plays a relatively minor role in our investigation of very wide companions.

\subsection{Comoving companion search}

We search for very wide comoving companions with projected separations up to $10^6$\,AU ($4.8$\,pc). We start with a selection of nearby stars, from which we further select the comoving companions. For each target star (hot jupiter host, contact binary, or field star), we select its nearby stars where (1) either the parallax difference $< 0.2$\,mas or the difference of line-of-sight distance (inverse of parallax) is $<20$\,pc; (2) projected physical separations $<20$\,pc assuming all nearby stars have the same parallaxes as the target star. Furthermore, we require that the candidate comoving companion meet all criteria in Sec.~\ref{sec:gaia}, except that we do not apply criteria on Gaia DR2 parameters {\tt phot\_rp\_mean\_flux\_over\_error}, {\tt phot\_bp\_mean\_flux\_over\_error}, {\tt phot\_bp\_rp\_excess\_factor}, BP$-$RP, and $\Delta \rm{G}$ since these criteria may exclude faint companions (like M dwarfs) and we are interested in identifying comoving companions of all stellar types.

%(3) other basic criteria explained in Sec.~\ref{sec:gaia}, except that we do not apply criteria on {\tt phot\_rp\_mean\_flux\_over\_error}, {\tt phot\_bp\_mean\_flux\_over\_error}, and {\tt phot\_bp\_rp\_excess\_factor} because these selections may exclude faint companions (like M-dwarfs) and the color of the companion is not the main focus in this paper. 

For the nearby stars selected following this procedure, we then compute the projected relative velocity between two stars using their proper motion difference and their mean parallax. The projected physical separations are recomputed but now using the mean parallax of two stars. We use projected physical separations instead of 3-D physical separations because the 3-D physical separations are dominated by the parallax uncertainty.

We remove targets (hot jupiter hosts, contact binaries, and their matched field stars) that are in clusters or comoving groups, because they may contaminate the comoving companion search. Specifically, we remove target stars that have $\ge 100$ nearby stars within the separations of $10^{5-6}$\,AU and relative velocities of $<10$\kms. This only removes $\lesssim1$\% of the samples and therefore does not affect the main results. 
%\blue{I'll need to explain in detail that, no multiple comoving companions, bla bla bla. And how I make sure that they all have comoving...}

%This only removes 3 out of 2311 contact binaries, 1 out of their matched field stars, 3 out of 262 hot jupiter hosts, and 4 out the their matched field stars. Only $\lesssim1$\% of the sample are removed and therefore it does not affect our results. 

%Different from other samples, the wide comoving companions around hot jupiter hosts in Fig.~\ref{fig:separation-dist} start to increase when separations $>10^5$\,AU. This is due to the contamination from the comoving groups (Fig.~\ref{fig:HJ-comoving}), and this contamination is stronger for hot jupiter hosts probably because they are younger \citep{Hamer2019}. Therefore, we exclude separations $>10^5$\,AU to avoid the comoving groups \red{(maybe these are dominated by a few hot jupiters, not all of them? Maybe I want to exclude them.)}. 

%While Gaia's theoretical spatial resolution is $0$\farcs{12}, current Gaia DR2 does not resolve every object pairs with separations up to $\sim0$\farcs{5} (e.g. \citealt{Arenou2018,Hwang2019a}), which might be due to the scanning direction or limitations of resolving small-separation pairs with high flux contrast. Our contact binaries are limited to parallax $>0.5$\,

\begin{figure*}
	\centering
	\includegraphics[height=.34\linewidth]{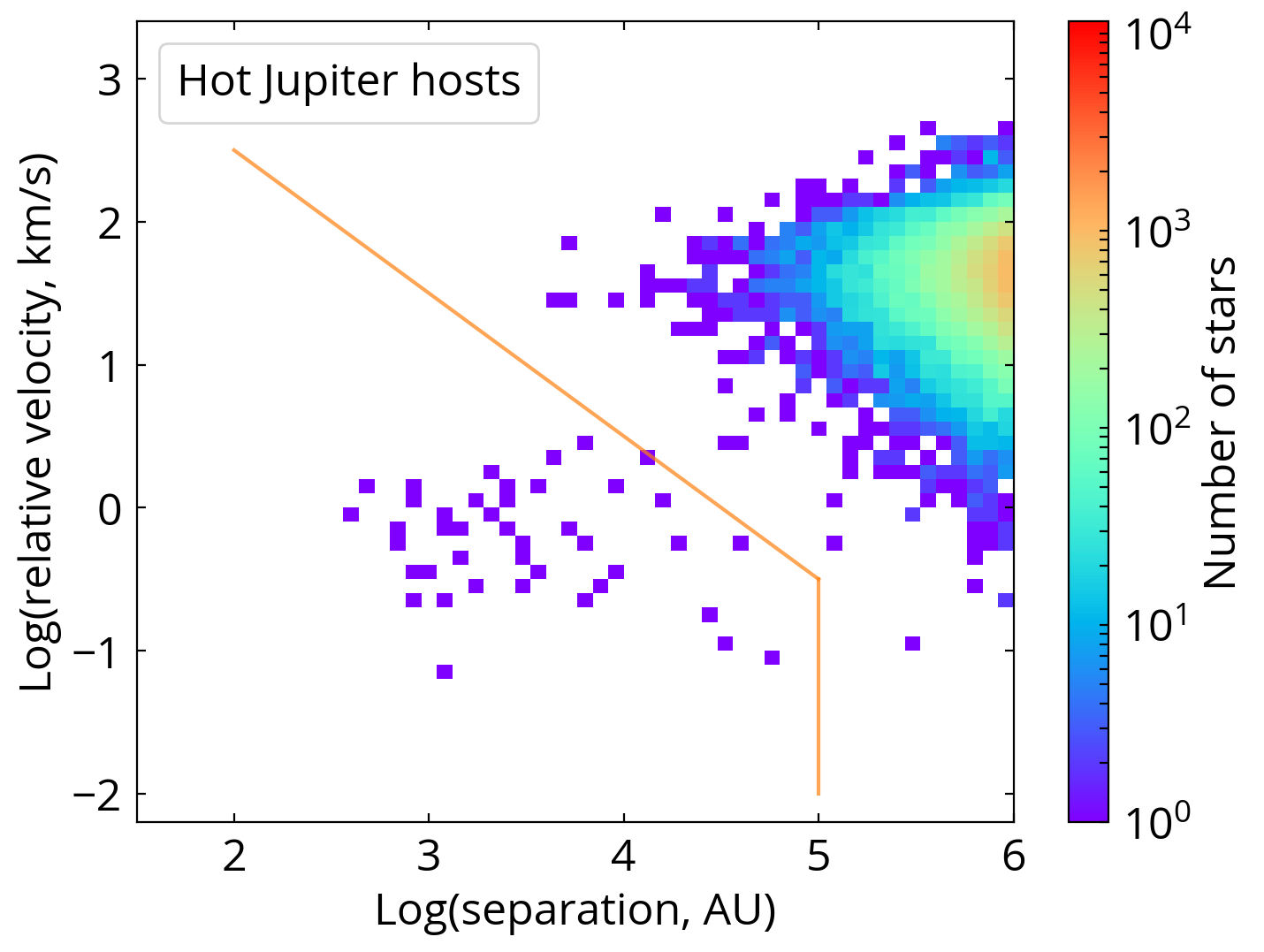}
	\includegraphics[height=.34\linewidth]{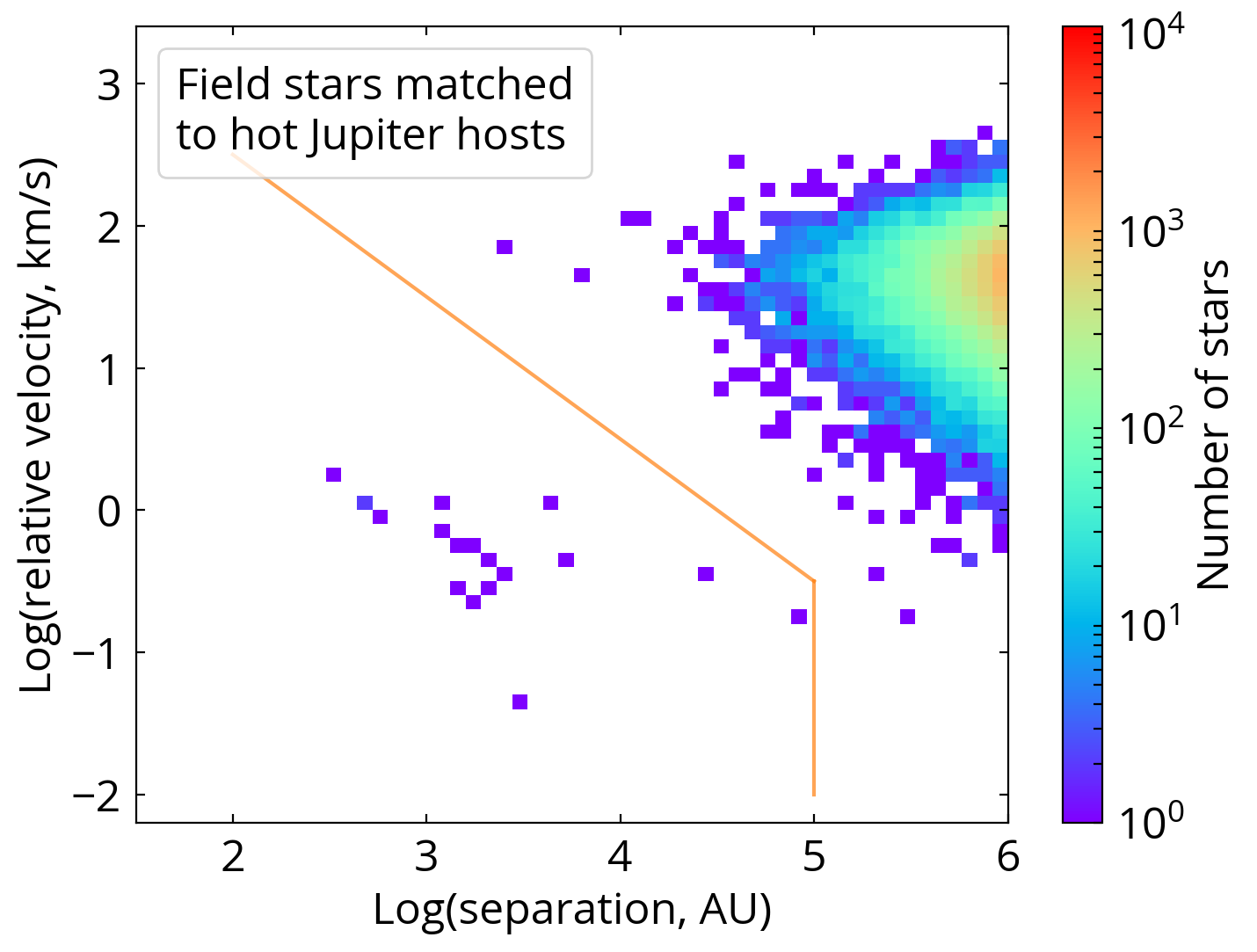}
	\caption{Search of comoving companions around hot jupiter hosts (left) and their matched field stars (right). The x-axis is the physical separations projected on the sky assuming the mean of the two parallaxes, and the y-axis is the projected relative velocity. The upper-right distribution corresponds to the physically unrelated nearby stars, while the lower-left is the comoving companions. The solid line is the empirical demarcation line to isolate the comoving companions. Although 4 matched field stars are used for each hot jupiter host, here we only show the same number of field stars and hot jupiter hosts for better comparison. }
	\label{fig:HJ-comoving}
\end{figure*}

\begin{figure*}
	\centering
	\includegraphics[height=.34\linewidth]{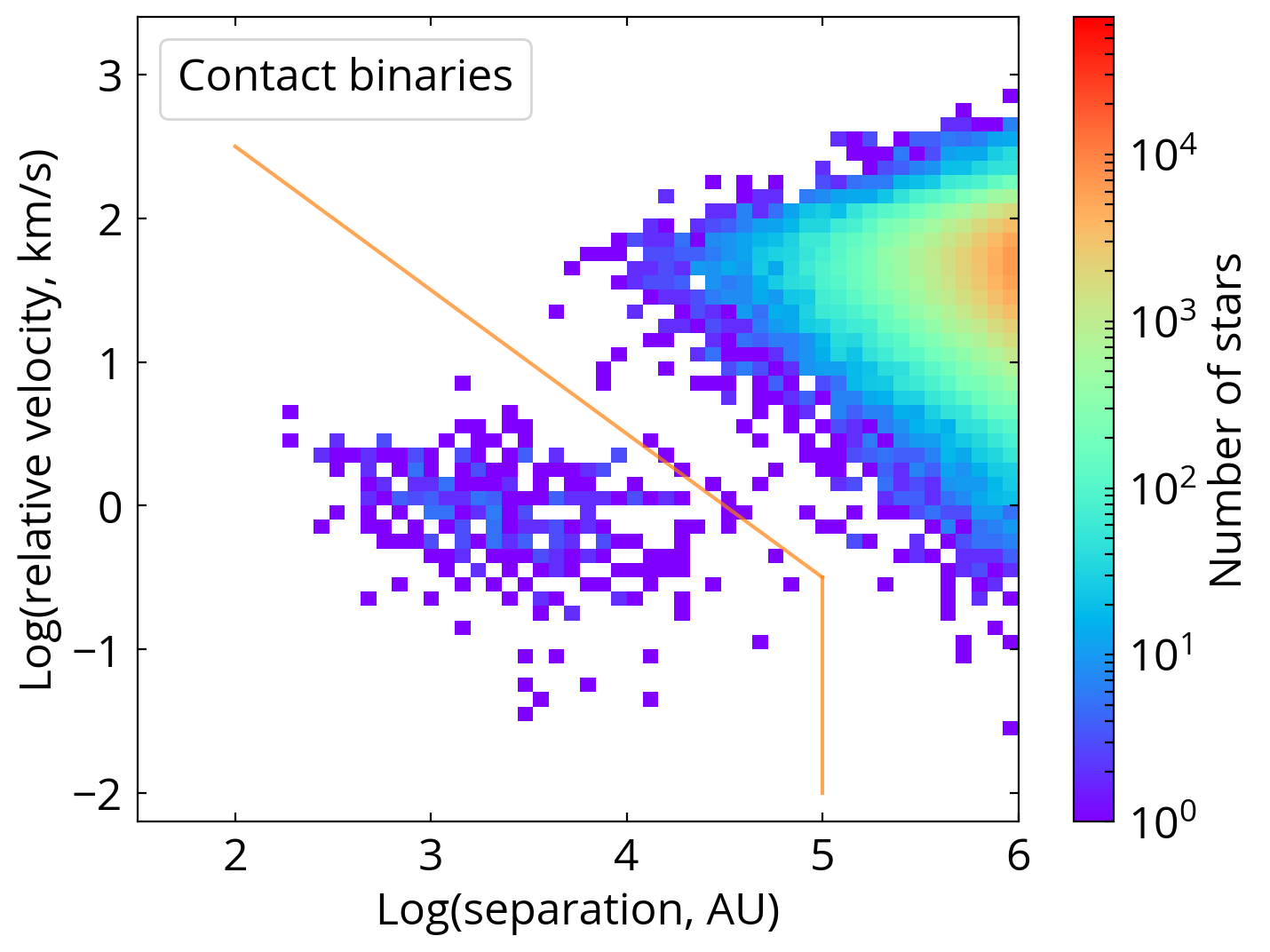}
	\includegraphics[height=.34\linewidth]{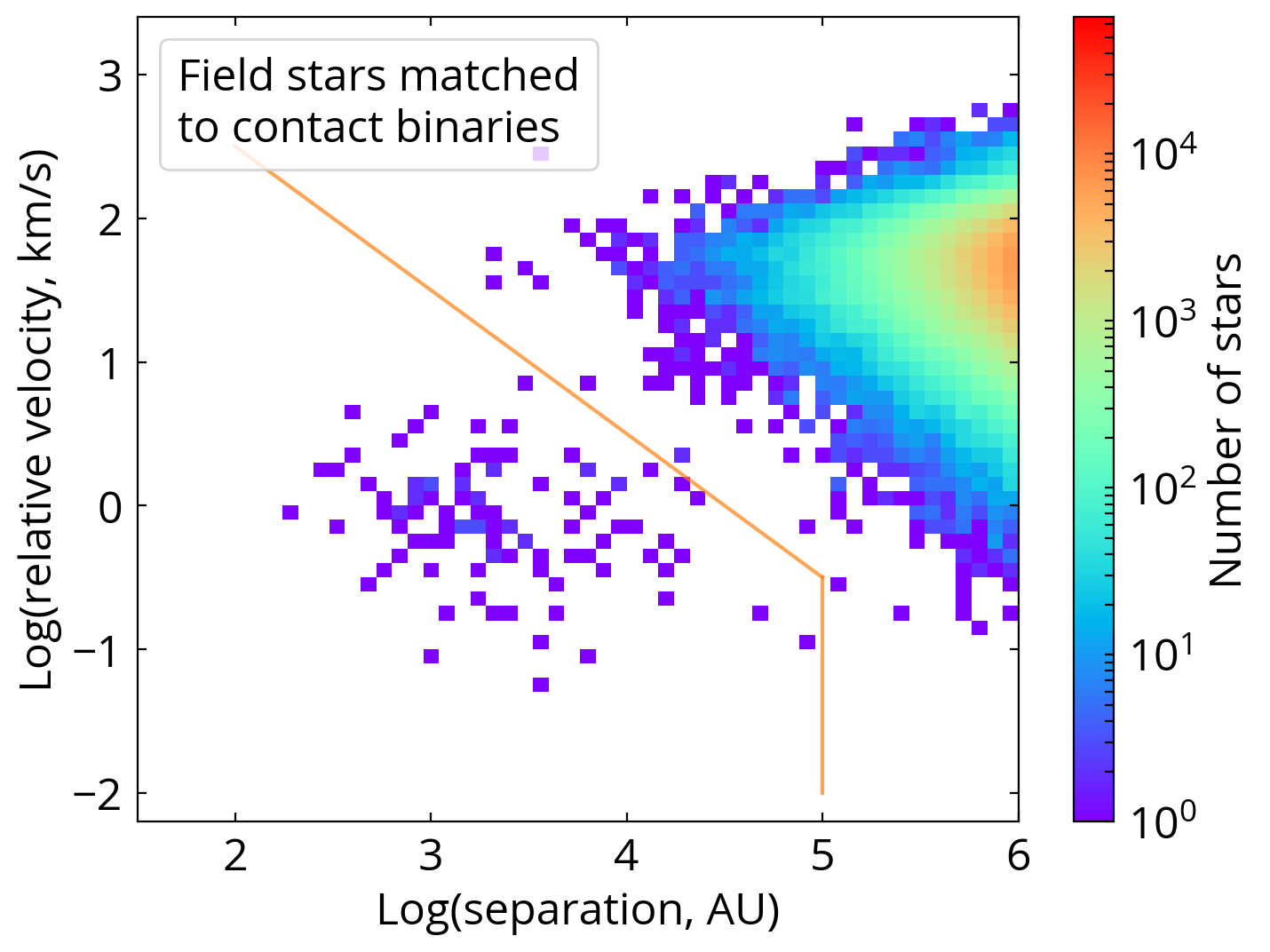}
	\caption{Similar to Fig.~\ref{fig:HJ-comoving}, but here for main-sequence contact binaries (left) and their matched field stars (right). Every contact binary has one matched field star.}
	\label{fig:EB-comoving}
\end{figure*}

\begin{figure*}
	\centering
	\includegraphics[height=.34\linewidth]{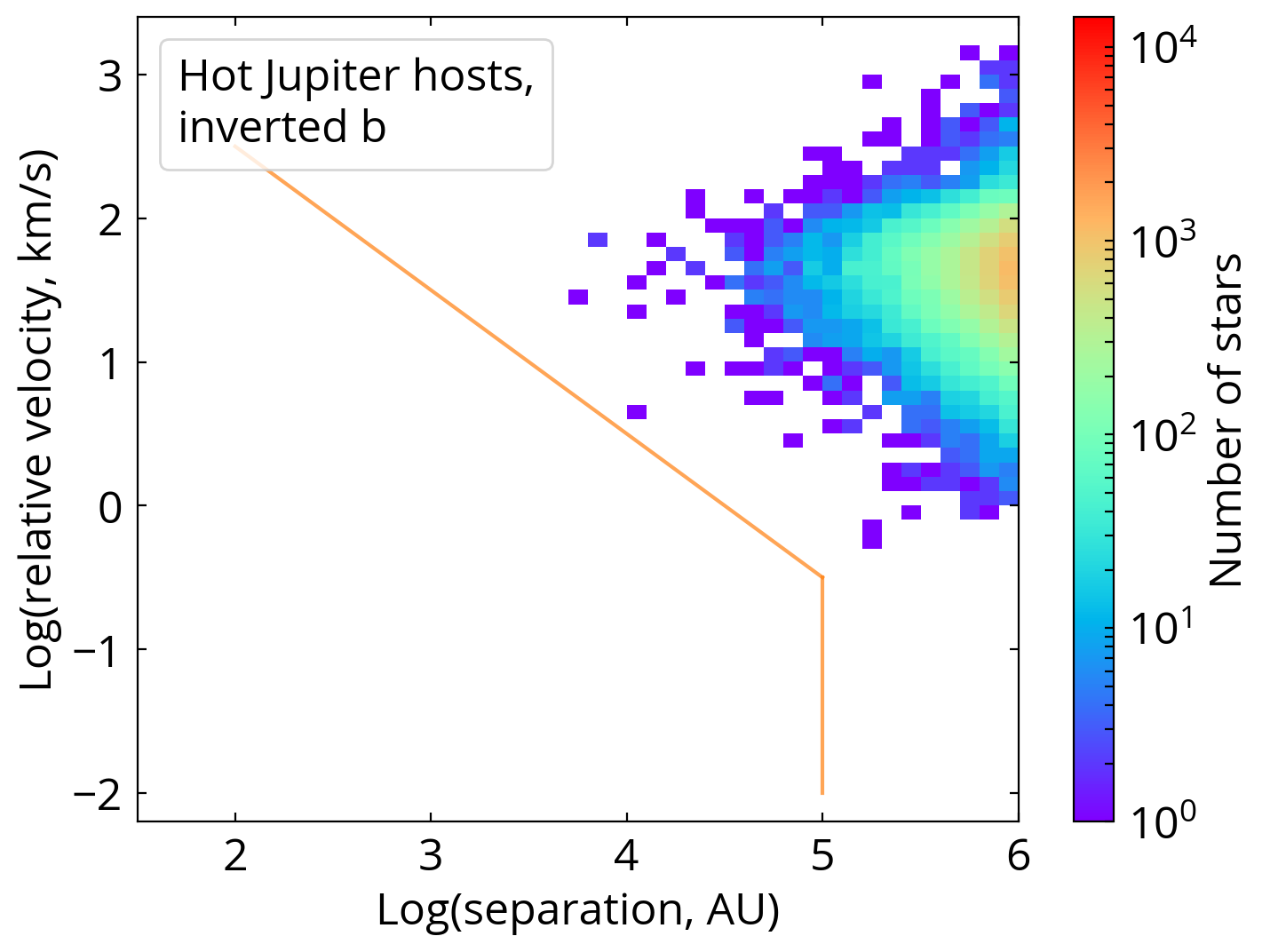}
	\includegraphics[height=.34\linewidth]{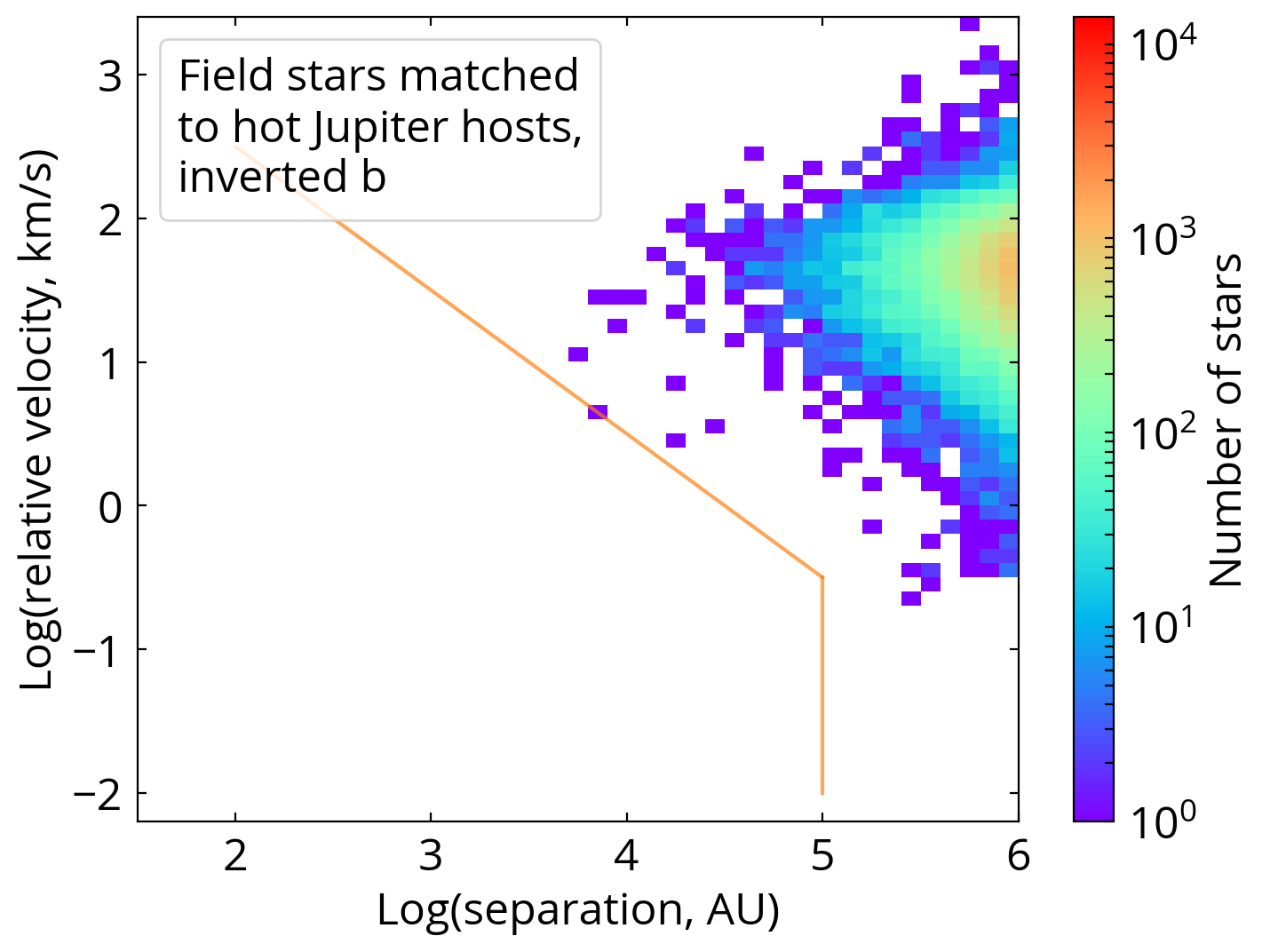}
	\caption{Contamination tests for hot jupiter hosts (left) and their matched field stars (right) by inverting their Galactic latitudes. The axes are the same as Fig.~\ref{fig:EB-comoving}. Although 4 matched field stars are used for each hot jupiter host, here we only show the same number of field stars and hot jupiter hosts for better comparison. }
	\label{fig:HJ-comoving-ib}
\end{figure*}

\begin{figure*}
	\centering
	\includegraphics[height=.34\linewidth]{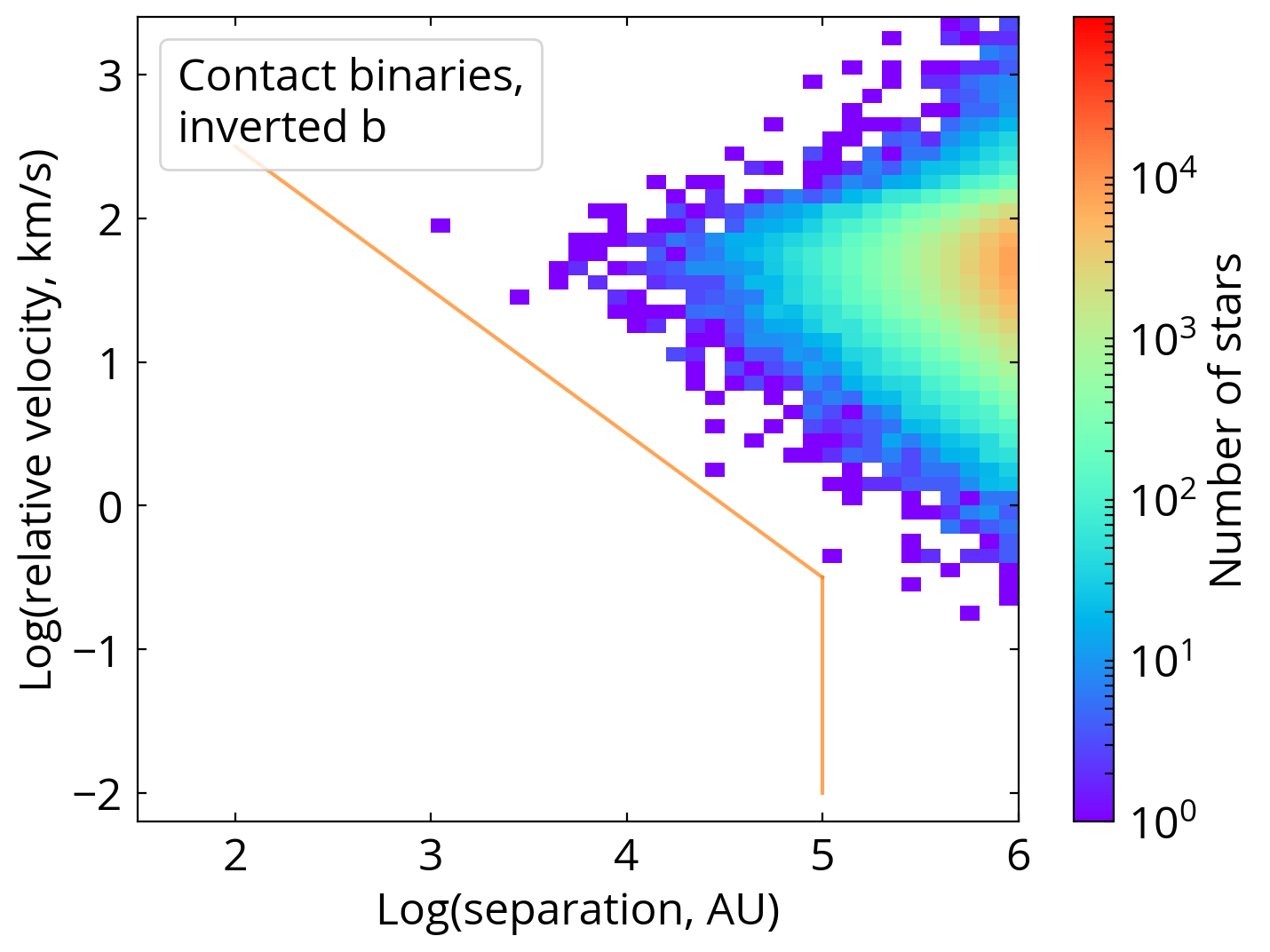}
	\includegraphics[height=.34\linewidth]{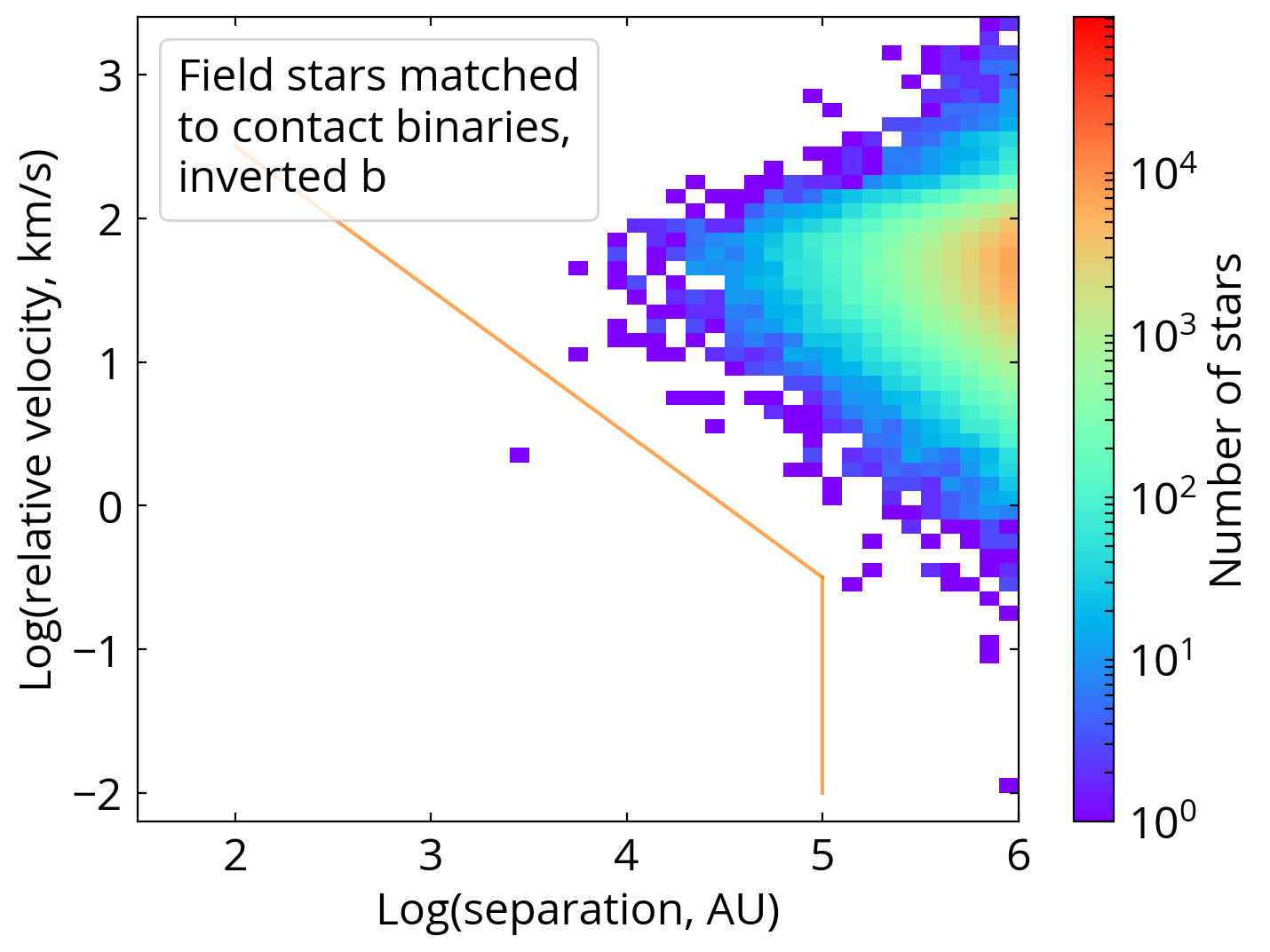}
	\caption{Contamination tests for main-sequence contact binaries (left) and their matched field stars (right) by inverting their Galactic latitudes. }
	\label{fig:EB-comoving-ib}
\end{figure*}

\begin{figure}
	\centering
	\includegraphics[width=.9\linewidth]{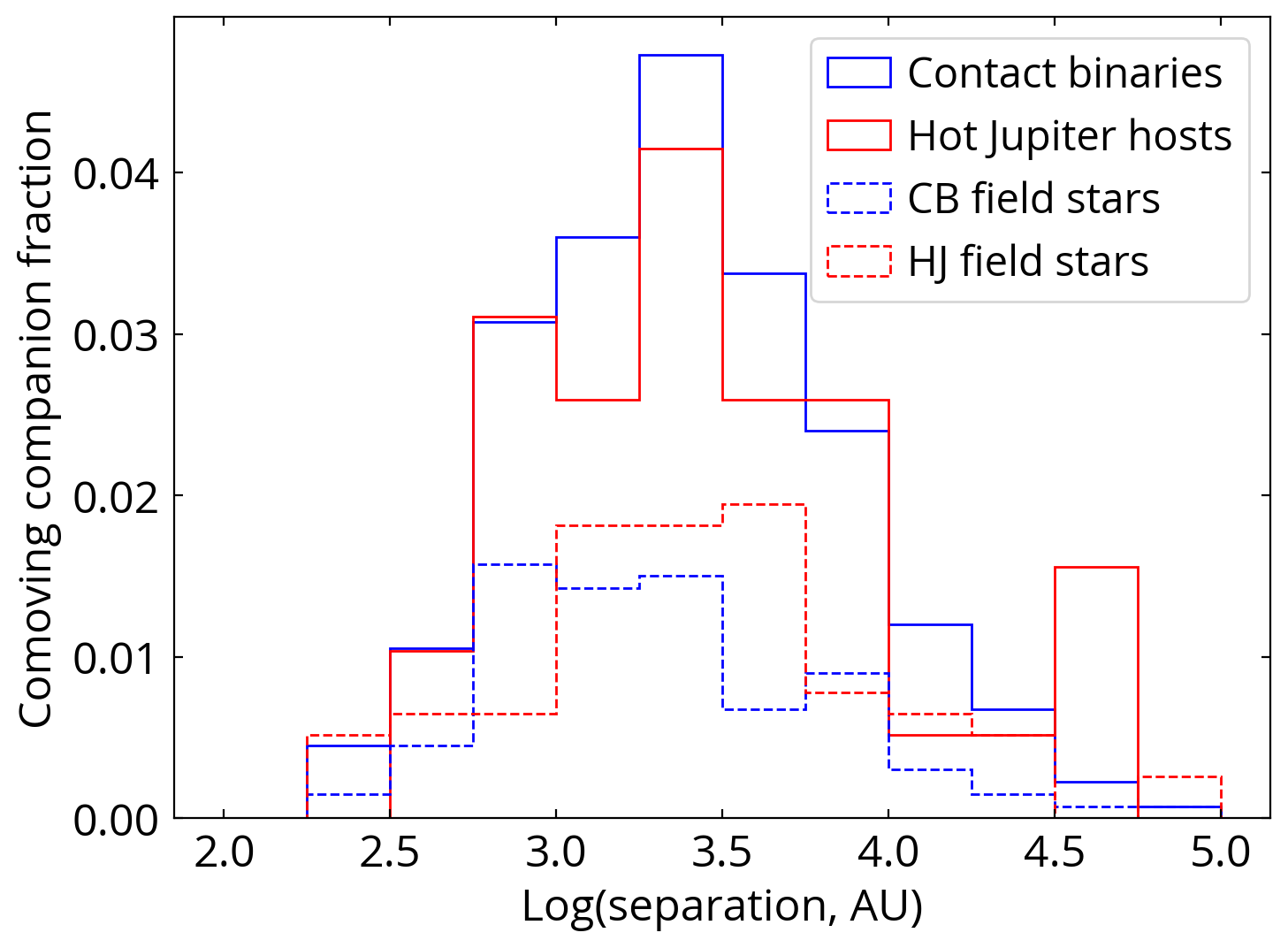}
	\caption{The distributions of projected separations for the comoving companions around hot jupiter hosts, main-sequence contact binaries, and their matched field stars. For separations $<10^3$\,AU, the sample starts to suffer from incompleteness due to Gaia photometry. The separations of $10^3-10^4$\,AU is secure for the investigation of wide comoving companions.}
	\label{fig:separation-dist}
\end{figure}

%In principle we can compute 3-D physical separations by taking the parallax differences into account, but the error of 3-D physical separations from the parallax uncertainty is .

\subsection{Selection of very wide comoving companions}

Fig.~\ref{fig:HJ-comoving} and Fig.~\ref{fig:EB-comoving} show the distributions of relative velocity versus projected physical separation of nearby sources for hot jupiter hosts and contact binaries, respectively. In these plots, the over-density at the upper-right corner is the chance projection stars, and the over-density at the lower-left corner is the comoving companions of the target stars. The triangular shape of the chance projections is because in the log-log space of relative velocity and physical separation, the number of chance projections increases as a power of 2 with respect to separation. The number of chance projections increases as a power of 2 with respect to the velocity in the low velocity limit of the Maxwellian-like velocity distribution (the velocity difference in Fig.~\ref{fig:HJ-comoving} and Fig.~\ref{fig:EB-comoving} contains two dimensions of velocity). At the high velocity limit of the Maxwellian-like distribution, the number of chance projections decreases exponentially. Therefore, the constant number of chance projection follows a slope of $-1$ at low velocities. 

%\red{(Are you going to show your beautiful plot that you obtain by flipping the sign of b to demonstrate that these are chance projections?.. Also, can you explain why the chance projections cloud has the shape that it does? Is this is mainly the Gaia selection effect, and if so, which one? In Figure 1, the chance projection cloud is slightly different in shape on the left than on the right. )}

The orange solid line in Fig.~\ref{fig:HJ-comoving} and Fig.~\ref{fig:EB-comoving} is an empirical demarcation line that isolates the comoving companions from the chance projection stars. Specifically, this demarcation line has a relative velocity of $10^{1.5}$\kms\ at a projected separation of $10^3$\,AU and a slope of $-1$ in the log-log space, and removes objects beyond $10^5$\,AU.  The slope of the demarcation line ensures that it is parallel to the chance projection stars so that the contamination level is not a strong function of the separation. 
%\red{(I like this... But would be even better if we had some understanding of why it's -1, see above...)}

%For a binary of two solar-mass stars, their Keplerian non-projected orbital velocity is 0.67\kms\ at $10^3$\,AU, and the orbital velocity decreases as a power of $-0.5$ with respect to the separation. The slope of our demarcation line is to ensure that it is parallel to the random physically unrelated stars, so that the contamination is not a strong function as the separation. 

%For random physically unrelated stars, their number density in the log-log space of separation and relative velocity increases as a power of 2 with respect to separation, and increases as a power of 3 with respect to velocity assuming a Maxwellian-like 1-D velocity distribution at the low-velocity end \red{(???)}.

Given the Gaia DR2 sensitivity down to $\sim20$\,mag in G-band, our search for comoving companions is complete down to an absolute G-band magnitude of 12\,mag within 400 pc. When selecting comoving companions, we adopt a conservative cut that the comoving companions have G-band absolute magnitudes brighter than 11.5\,mag. Therefore, we should detect most of the stellar objects, except for late M dwarfs and old white dwarfs. Since the hot jupiter hosts in wide binaries and wide solar-type binaries have statistically consistent stellar mass-ratio distributions \citep{Moe2019b}, the incompleteness at the faintest end does not affect our result.

In Fig.~\ref{fig:HJ-comoving-ib} and Fig.~\ref{fig:EB-comoving-ib}, we test the level of contamination from the random background stars by flipping the sign of the Galactic latitudes of the targets \citep{Shaya2011, Jimenez-Esteban2019}. In addition, we flip the sign of the proper motion in the direction of galactic latitudes after removing the solar motion and the differential rotation from the Galactic disk. The tests show that 0, 1, 0, 1 chance projection stars fall into our selection (below the orange demarcation line and a separation between $10^3-10^4$\,AU) for hot jupiter hosts, the field stars matched for hot jupiter hosts, contact binaries, and the field stars matched for the contact binaries, respectively. The contamination from the chance projection stars is mainly due to the targets at low Galactic latitudes. This level of contamination does not significantly affect our main results. 

Fig.~\ref{fig:separation-dist} shows the separation distribution of the comoving companions selected using the demarcation line. The comoving companion fraction on the vertical axis is computed from the number of comoving companions in each separation bin divided by the total number of the sample. We remove targets (field stars) and their corresponding field stars (targets) if more than one comoving companions are found. Specifically, none of hot jupiter hosts and the field star samples has 2 resolved comoving stars. In contrast, a significantly higher fraction (7/1333) of contact binaries have 2 comoving stars, providing further constraints on the formation of contact binaries and their wide companions (Hwang et al. in preparation). Overall, multiple comoving systems are rare so removing them does not affect the results. 

%7 ($\sim0.5$\%) contact binaries have 2 comoving stars, significantly higher than ($\sim1\%$ for the hot jupiter hosts and the contact binaries) 

We find one known dual hot jupiter host, WASP-94 A and B \citep{Neveu-Vanmalle2014,Teske2016}, and one known dual contact binary, BV Dra and BW Dra \citep{Batten1965}, and one newly discovered dual binary. In our procedure, we count the separations of these dual systems twice, but it does not change the main result if we exclude them or count their separations once.

We provide two machine-readable tables, one for the hot jupiter hosts and the other one for the contact binaries, with their corresponding wide companions that are used in Fig.~\ref{fig:separation-dist}. The tables contain (1) the Gaia DR2 \texttt{source\_id} of hot jupiter hosts and contact binaries; (2) the Gaia DR2 \texttt{source\_id} of their companions; (3) their physical separations in AU. Dual systems appear two times in the table.

Since we require reliable BP$-$RP colors for the targets (not for the companions), our spatial resolution is limited by Gaia's BP- and RP-band photometry because Gaia DR2 uses a window of $3.5\times2.1$\,arcsec$^2$ to measure the total flux in BP- and RP-bands, i.e. applies no deblending. For pairs with separations $\lesssim 2$\,arcsec, their BP and RP fluxes may be affected by the companion and may be excluded by the {\tt phot\_bp\_rp\_excess\_factor} criteria. Therefore, our secure spatial resolution is 2\,arcsec, and because we limit our sample to be within 400\,pc, we are able to probe the very wide comoving companions with separations $>800$\,AU.

The contact binary sample and two field star samples all show a steady decline of the companion fractions toward larger separations, but the hot jupiter sample has an enhanced comoving companion fraction at $\sim10^{4.5}$\,AU. While it may be due to the contamination from chance projection, the contamination test (Fig.~\ref{fig:HJ-comoving-ib}) shows that $<1$ contamination is expected at this separation. However, there are only three sources in the bin at $\sim10^{4.5}$\,AU, and therefore it suffers from small-number statistics and its significance requires a larger sample to confirm. 

%Therefore, this enhancement at  $\sim10^{4.5}$\,AU might be real. 

%These companions at $\sim10^{4.5}$\,AU may be the wide binaries that are slowly drifting apart \red{(due to what mechanism? )} \citep{Oh2017}. Furthermore, the separation of $10^{4.5}\sim30,000$\,AU coincides with the theoretical prediction at $\sim40,000$\,pc for solar-mass binaries \red{(theoretical prediction of what? )} \citep{Jiang2010a}.
 
We conclude that separations of $10^3-10^4$\,AU is the secure separation range to investigate in Gaia DR2, and we define the very wide companion fraction as the fraction of a sample that have comoving companions at separations of $10^3-10^4$\,AU. In the next section, we investigate the very wide comoving fraction in hot jupiter hosts and contact binaries.
 
%or small comoving groups that have not dissolved in the Galaxy completely \red{(See the Oh et al paper I sent around)}. Therefore, we conclude that separations of $10^3-10^4$\,AU is the secure separation range to investigate the wide comoving companions fraction in Gaia DR2.

%and the hot jupiter hosts and the contact binary sample show enhanced comoving fractions compared to their field star samples. 

%In the following sections, we use $F_{\rm HJ,wide}$ for the fraction of hot jupiter hosts that have comoving companions with separations of $10^{3-4}$\,AU. Similarly, we use $F_{\rm CB,wide}$ for contact binaries, and $F_{\rm HJfield,wide}$ and $F_{\rm CBfield,wide}$ for their matched field samples. Then we compute the wide companion enhancement factor by $f_{\rm enhance,HJ}=F_{\rm HJ,wide}/F_{\rm HJfield,wide}$ and $f_{\rm enhance,CB}=F_{\rm CB,wide}/F_{\rm CBfield,wide}$. 

\section{Results}
\label{sec:result}

\begin{table*}[]
	\label{tab:result}
	\caption{Very wide companion fraction ($10^3-10^4$\,AU) for hot jupiter hosts and contact binaries.}
	\center
	\begin{tabular}{ccc}
\hline \hline
		 & Hot jupiter hosts & Contact binaries \\
\hline
		Very wide companion fraction ($10^3-10^4$\,AU)                                      & 11.9$\pm$2.5\% (23/193)           & 14.1$\pm$1.0\% (188/1333)         \\
		Field star values                                                      			&  6.3$\pm$0.9\% (49/772)            & 4.5$\pm$0.6\% (60/1333)          \\
		Enhancement compared to the field values                      	&  1.9$\pm$0.5                 	 		 	&       3.1$\pm$0.5           \\
		Enhancement expected from the lack of close companions &  $1.4\pm0.3$							                 &     $1.0$    \\   
\hline \hline     
	\end{tabular}
\end{table*}

%Fig.~\ref{fig:enhancement} shows that both hot jupiter hosts and main sequence contact binaries have higher chances of having a wide companions ($10^{3-4}$\,AU) compared to their matched field star samples. Specifically, $F_{\rm HJ,wide}=10.6\pm 2.2$\% (23/217) of hot jupiter hosts have companions in separations of $10^{3-4}$\,AU, while only $F_{\rm HJfield,wide}=4.1\pm 0.7$\% (35/864, \blue{I need to check why is 864, not $217*4=868$}) for their matched field sample. Therefore, hot jupiter hosts have an enhanced wide comoving companions by a factor of $f_{\rm enhance,HJ}=F_{\rm HJ,wide}/F_{\rm HJfield,wide} = 2.6\pm 0.7$. For main sequence contact binaries, $13.1\pm 0.8$\% (301/2293) have such wide companions, significantly higher than $4.4\pm 0.4$\% (100/2293) in their matched field star sample, with an enhancemant factor of $f_{\rm enhance,CB}=F_{\rm CB,wide}/F_{\rm CBfield,wide} = 3.0\pm0.3$.

In this section, we start with the presentation of the observed very wide companion fractions in Sec.~\ref{sec:result-observed}. We then compute how the lack of close companions may affect the very wide companion fractions in Sec.~\ref{sec:result-expected}. We compare the observed quantities with the enhancement due to the lack of close companions in Sec.~\ref{sec:result-comparison}.

\subsection{Observed very wide companion fractions}
\label{sec:result-observed}

Table~\ref{tab:result} presents the very wide comoving companion fractions for hot jupiter hosts, contact binaries, and their matched field stars. Specifically, 11.9$\pm$2.5\% (23/193) of hot jupiter hosts and 6.3$\pm$0.9\% (49/772) of their matched field stars have companions at separations of $10^3-10^4$\,AU. For main-sequence contact binaries and their matched field stars, 14.1$\pm$1.0\% (188/1333) and 4.5$\pm$0.6\% (60/1333) have very wide comoving companions. The ratios of the observed comoving fraction to the field value is 1.9$\pm$0.5 and 3.1$\pm$0.5 for the hot jupiter hosts and main-sequence contact binaries, respectively.

%Fig.~\ref{fig:enhancement} shows that both hot jupiter hosts and main sequence contact binaries have higher chances of having a wide companions ($10^{3-4}$\,AU) compared to their matched field star samples. Specifically, $9.0\pm 1.6$\% (31/346) of hot jupiter hosts have companions in separations of $10^{3-4}$\,AU, while only $3.8\pm 1.2$\% for their matched field sample. For main sequence contact binaries, $11.6\pm 0.5$\% (553/4781) have such wide companions, significantly higher than $3.5\pm 0.3$\% from their matched field star sample. 

%\blue{talk about the completeness of companions...}

%\blue{Given the Gaia DR2 sensitivity down} to $\sim20.5$\,mag in G-band, our search for comoving companions is complete down to an absolute G-band magnitude of 12\,mag within 500 pc. Therefore, we should detect most of the stellar objects, except for late M dwarfs and old white dwarfs. We re-do the companion search \blue{(I don't like this)} with a parallax cut of 5\,mas (distances $<200$\,pc) so that the companion sample is complete down to absolute G-band magnitude of 14\,mag, and the results are similar. Since the hot jupiter hosts and wide solar-type binaries have mutually, statistically consistent mass-ratio distributions \citep{Moe2019b}, the incompleteness at the faint end does not affect our result.

%\subsection{Accounting for the selection effect}

\subsection{Expected enhancement of very wide companion fraction due to the lack of close companions}
\label{sec:result-expected}

\cite{Moe2019b} point out that the seemingly enhanced wide companion fraction of hot jupiter hosts compared to the field stars may be the consequence of the lack of close companions ($<50$\,AU) to the hot jupiter hosts. As an extreme example, if all the stars were in stellar binary systems with a wide range of separations and hot jupiters could not form in binaries with separations $<1000$\,AU, then we would find that all hot jupiter hosts have wide stellar companions at separations $>1000$\,AU, i.e. an enhanced very wide companion fraction compared to the field stars. Therefore, an ideal comparison sample of the field stars for the hot jupiter hosts would be those field stars that do not have close companions within $\sim50$\,AU. Unfortunately, it is not possible to select such a sample at the present time: such stellar pairs are too close to be spatially resolved, photometric selection on the color-magnitude diagram is rarely precise enough to distinguish a single from a binary with a high mass ratio, and the use of high-precision radial velocity surveys strongly reduce the sample size. Therefore, there is no way to evaluate the fraction that have a companion within $\sim50$\,AU in our field star samples.

%We begin with the definition of our notations. Given a sample, say hot jupiter hosts, we consider their single, binary, and triple systems. Higher-order multiples are neglected here because they are rare, and their separation is not well defined, depending on the choice of the star at the origin.

As we are unable to find comparison objects without close companions, we test the hypothesis that the hot jupiter hosts are a random sampling of the field stars except that hot jupiter hosts avoid systems with close companions. If there were no triple systems, this statement is equivalent to having the same the companion separation distribution for hot jupiter hosts and for the field stars at separations $\gtrsim50$\,AU. If there are triples, the companion separation distributions of hot jupiter hosts and field stars are not the same, because avoiding systems with close separations also affects the companions at large separations.

Under such hypothesis, we compute the expected enhancement of companion fraction at separations between $s_0$ and $s_1$ due to the lack of close companions, denoted as $\mathcal{E}^{\rm HJ}(s_0,\ s_1)$:

\begin{equation}
\label{eq:1}
	\begin{multlined}
		\mathcal{E}^{\rm HJ}(s_0,\ s_1) = \\ \frac{F^{\rm HJ}(s_0,\ s_1)}{F^{\rm field}(s_0,\ s_1)} 
		=\frac{B^{\rm HJ}(s_0,\ s_1) + T^{\rm HJ}(s_0,\ s_1) }{B^{\rm field}(s_0,\ s_1) + T^{\rm field}(s_0,\ s_1)},
	\end{multlined}
\end{equation}
where $F^{\rm HJ}(s_0,\ s_1)$ is the companion fraction of hot jupiter hosts at separations between $s_0$ and $s_1$; $B^{\rm HJ}(s_0,\ s_1)$ is the fraction of hot jupiter hosts that are in binary systems with separations between $s_0$ and $s_1$; and $T^{\rm HJ}(s_0,\ s_1)$ is the companion fraction with separations between $s_0$ and $s_1$ contributed by triple systems. For field stars, similar definitions are used for $F^{\rm field}(s_0,\ s_1)$, $B^{\rm field}(s_0,\ s_1)$, and $T^{\rm field}(s_0,\ s_1)$.
%where $F_{\rm HJ}(s_0,\ s_1)$ is the companion fraction of hot jupiter hosts at separations between $s_0$ and $s_1$; $B_{\rm HJ}(s_0,\ s_1)$ is the fraction of hot jupiter hosts that are in binary systems with separations between $s_0$ and $s_1$; $T_{\rm HJ,in}(s_0,\ s_1)$ is the fraction of hot jupiter hosts that are in the triple systems with inner separations between $s_0$ and $s_1$; $T_{\rm HJ,out}(s_0,\ s_1)$ is the fraction of hot jupiter hosts that are in the triple systems with outer separations between $s_0$ and $s_1$. For field stars, similar definitions are used for $F_{\rm field}(s_0,\ s_1)$, $B_{\rm field}(s_0,\ s_1)$, $T_{\rm field,in}(s_0,\ s_1)$, and $T_{\rm field,out}(s_0,\ s_1)$.

%For field stars, $T_{\rm field,in}(s_0,\ s_1) = 2 T_{\rm field,out}(s_0,\ s_1)$ because a triple consists of two stars in the inner binary and one outer star.

%$T_{\rm HJ,in}=T_{\rm HJ,out}=T_{\rm field, in}=T_{\rm field, out}=0$
We first calculate the expected enhancement of companion fraction ($\mathcal{E}^{\rm HJ}_{\rm no\ triple}$) in the case with no triples, i.e. $T^{\rm HJ}=T^{\rm field}=0$. We use {\it close} to denote the separations smaller than 50\,AU, {\it wide} to denote the separations between 50 and 2000\,AU, and {\it very\ wide} to denote the separations between $10^3$ and $10^4$\,AU. From Equation~\ref{eq:1}, we have $\mathcal{E}^{\rm HJ}_{\rm no\ triple}({wide})= B^{\rm HJ}({wide})/B^{\rm field}({wide})$. We further define a function $\mathcal{S}^{\rm HJ} (s_0, s_1)\equiv B^{\rm HJ} (s_0, s_1) / (1 - B^{\rm HJ}(close))$ and similarly for $\mathcal{S}^{\rm field} (s_0, s_1)$. Then under the hypothesis that the shapes of the companion separation distributions of hot jupiter hosts and field stars are the same beyond 50\,AU, we have $\mathcal{S}^{\rm HJ}({wide})=\mathcal{S}^{\rm field}({wide})$. Therefore, $\mathcal{E}^{\rm HJ}_{\rm no\ triple}({wide}) = B^{\rm HJ}({wide})/B^{\rm field}({wide})=[\mathcal{S}^{\rm HJ}({wide})(1-B^{\rm HJ}({close}))]/[\mathcal{S}^{\rm field}({wide})(1-B^{\rm field}({close}))]=(1-B^{\rm HJ}({close}))/(1-B^ {\rm field}({close}))$. With $F^{\rm HJ}(close)=B^{\rm HJ}(close)=0.04$ \citep{Ngo2016} and $F^{\rm field}({close})=B^{\rm field}(close)=0.40$ \citep{Moe2019b}, the expected enhancement for very wide companion fractions is $\mathcal{E}^{\rm HJ}_{\rm no\ triple}({wide}) = (1-0.04)/(1-0.40) = 1.6$. Thus in the case of no triples, there is an apparent enhancement of companion fraction which is independent of the chosen separation range once it is $>50$\,AU, and therefore $\mathcal{E}^{\rm HJ}_{\rm no\ triple}({\rm wide}) = \mathcal{E}^{\rm HJ}_{\rm no\ triple}({\rm very\ wide})$. This is an example that the (very) wide companion fraction of hot jupiter hosts may be enhanced compared to that of the field star sample due to the lack of close companions in hot jupiter hosts which are common in the comparison sample of field stars.

The contribution of companion fraction from triples can be written as
\begin{equation}
\begin{multlined}
T^{\rm HJ}(s_0,\ s_1) = \\ T^{\rm HJ}_{\rm in,0}(s_0,\ s_1) + T^{\rm HJ}_{\rm in,1}(s_0,\ s_1) + T^{\rm HJ}_{\rm out,1}(s_0,\ s_1),
\end{multlined}
\end{equation}
where $T^{\rm HJ}_{\rm in,0}(s_0,\ s_1)$ is the fraction of hot jupiter hosts that are in the inner binary of triples with inner separations between $s_0$ and $s_1$; $T^{\rm HJ}_{\rm in,1}(s_0,\ s_1)$ is the fraction of hot jupiter hosts that are in the inner binary of triples with outer separations between $s_0$ and $s_1$; and $T^{\rm HJ}_{\rm out,1}(s_0,\ s_1)$ is the fraction of hot jupiter hosts that are in the outer tertiary of triples with outer separations between $s_0$ and $s_1$. Similar definitions apply to $T^{\rm field}_{\rm in,0}(s_0,\ s_1)$, $T^{\rm field}_{\rm in,1}(s_0,\ s_1)$, and $T^{\rm field}_{\rm out,1}(s_0,\ s_1)$ for field stars.

%\begin{equation}
%\begin{multlined}
%\mathcal{E}(s_0,\ s_1) = \frac{F_{\rm HJ}(s_0,\ s_1)}{F_{\rm field}(s_0,\ s_1)} \\
%=\frac{B_{\rm HJ}(s_0,\ s_1) + T_{\rm HJ,in}(s_0,\ s_1) + T_{\rm HJ,out}(s_0,\ s_1)}{B_{\rm field}(s_0,\ s_1) + T_{\rm field,in}(s_0,\ %s_1) + T_{\rm field,out}(s_0,\ s_1)},
%\end{multlined}
%\end{equation}

Triples tend to make the enhancement of companion fraction ($\mathcal{E}$) smaller than the case without triples. For example, if there is a triple with inner separation $<50$\,AU and an outer separation of 5000\,AU, then this system would contribute a very wide companion to field stars through $T^{\rm field}_{\rm in,1}(very\ wide)$, but not to hot jupiter hosts through $T^{\rm HJ}_{\rm in,1}(very\ wide)$ because its small inner separation prevents the formation of a hot jupiter. An accurate estimate of $T^{\rm HJ}(s_0,\ s_1)$ is challenging because it requires a good understanding of the distributions of inner and outer separations and their correlation. 

By taking the contribution of triples into account, \cite{Moe2019b} estimate the expected enhancement of companion fraction due to the lack of close companions to be $\mathcal{E}^{\rm HJ}({\rm wide}) = 1.32\pm0.25$ ($0.37\pm0.07$/$0.28\pm0.05$) for wide binaries. For very wide companions investigated in this paper, $\mathcal{E}^{\rm HJ}({\rm very\ wide})$ may not be exactly the same as $\mathcal{E}^{\rm HJ}({\rm wide})$, depending on difference of the triple contribution in these two separation ranges. However, with current limited understanding of the correlation between inner and outer separations, we estimate that $\mathcal{E}^{\rm HJ}({\rm very\ wide})\sim \mathcal{E}^{\rm HJ}({\rm wide})$ to leading order.

%Even if the hot jupiter occurrence rate does not depend on their companion separation, the wide companion fraction of hot jupiter hosts may still be enhanced due to other selection effect. 

The unresolved inner binaries of triples may also enhance the (very) wide companion fractions. Unresolved binaries have two times higher probability of having a hot jupiter than single stars, simply because there are two stars in an unresolved binary. If there exists a significant number of triples with unresolved inner separations ($\lesssim 1000$\,AU in our case) and resolved outer separations ($> 1000$\,AU in our case), then hot jupiter hosts may have an enhanced wide companion fraction because they are more likely to be found in the inner binaries of triples. With some realistic binary fraction and triple fraction, we estimate that this effect would result in an enhancement of companion fraction of $10-20$\%, and may be smaller if hot jupiter hosts tend to have fewer close companions. Combined with the effect of the lack of close companions, the expected enhancement of very wide companion fraction is $\mathcal{E}^{\rm HJ}({\rm very\ wide}) = 1.4\pm0.3$ for hot jupiter hosts, and we tabulate this number in the bottom row of Table~\ref{tab:result}.

Unlike hot jupiter hosts where a lack of close companions is observed within $\sim$50\,AU, close binaries only show a lack of companions within $\sim1$\,AU \citep{Tokovinin2006, Gies2012,Tokovinin2014b}. Therefore, while the lack of close companions enhances the wide companion fraction for hot jupiter hosts, such effect is negligible for contact binaries \citep{Moe2019b}, and we tabulate $\mathcal{E}^{\rm CB}({\rm very\ wide}) = 1.0$ in the bottom row of Table~\ref{tab:result} for the contact binaries.

\subsection{A weak or no enhancement for hot jupiters, and a significant enhancement for contact binaries}
\label{sec:result-comparison}

Compared to $\mathcal{E}^{\rm HJ}({very\ wide}) =1.4\pm0.3$, our measured enhancement factor of $1.9\pm0.5$ for hot jupiter hosts suggests that there is no or a weak enhancement at $\sim1\sigma$ significance. Therefore, the enhanced very wide companion fraction from hot jupiter hosts is consistent, within $\sim1\sigma$, with the fact that they lack close companions. Confirming the weak enhancement requires a larger hot jupiter host sample in the future. 

Compared to $\mathcal{E}^{\rm CB}({very\ wide}) =1.0$, our measured enhancement factor of $3.1\pm0.5$ for main-sequence contact binaries shows that there is a significant enhancement at a $4\sigma$ significance. This enhancement cannot be explained by the lack of close companions, and therefore some other physical mechanisms are needed to explain the connection between contact binaries and their very wide companions.

% Contact binaries are also more likely to have two comoving companions than their matched field stars. Specifically, 7 out of 1113 tend to have 2 very wide companions, while none of the matched field stars have 2 very wide companions. 

\section{Discussion}

\label{sec:discussion}

\subsection{Very wide companions play a minor role in the orbital migration}

%Our results show that hot jupiter hosts have an enhanced very wide companion fraction compared to their matched field stars, and this enhancement can be explained by their lack of close companions. The enhanced very wide companion fraction of contact binaries cannot be explained by the same effect. 

Close binary fraction increases with decreasing metallicity, but (very) wide binary fraction with separations $\gtrsim200$\,AU is weakly dependent on the metallicity \citep{Moe2019, El-Badry2019a}. Therefore, the enhanced very wide companion fraction around contact binaries is not due to their metallicity.
% suggesting that there exists some physical reasons that enhance the formation of very wide companions around contact binaries. 

%The very wide companions at separations $>10^3$\,AU may be indicative about the formation mechanisms of proto-hot jupiters and proto-contact binaries, or about their orbital migration from larger separations to their current locations. However, 

The very wide companions investigated in this paper are not able to induce significant orbital migration in the proto-contact binaries and proto-hot jupiter systems through the classical Kozai-Lidov mechanism. For a companions at a separation of $10^3$\,AU, it can induce the Kozai-Lidov oscillation only when the inner binaries have a separation $\gtrsim5$\,AU, which is set by the requirement that the oscillation timescale needs to be shorter than the timescale of the relativistic pericenter precession \citep{Fabrycky2007}. Therefore, these very wide companions are not able to bring the inner proto-hot jupiter system and proto-contact binaries to their current separations ($<0.1$\,AU). 

When the outer orbit is eccentric, it can trigger a higher-order octuple effect, called the eccentric Kozai-Lidov effect \citep{Naoz2013,Naoz2016}. The eccentric Kozai-Lidov effect typically enhances the efficiency of forming hot jupiters \citep{Naoz2012} and close binaries \citep{Naoz2014}. Although the octuple timescale is not well quantified due to the chaotic nature of the eccentric Kozai-Lidov effect, it is typically longer than the classical Kozai-Lidov effect \citep{Naoz2016}. Therefore, the very wide companions are not able to contribute much to the orbital migration through the eccentric Kozai-Lidov effect either.

Another possibility is that the very wide companions were initially located at a smaller separation where they can induce strong Kozai-Lidov oscillations, and they migrated outward at a later time. This scenario is also suggested by \cite{El-Badry2019} where they find an excess of equal-mass stellar binaries out to separations of $\sim10^3$\,AU. Because equal-mass binaries are mainly formed from disk fragmentation at close separations ($\lesssim100$\,AU), they argue that the observed equal-mass wide binaries are formed with close separations and further widened by dynamical interactions in their birth environments. However, such excess of  equal-mass binaries at $\sim10^3$\,AU is only $\sim5$\% in the field. It is not yet clear if this outward migration is a dominant path for the very wide companions around contact binaries. Further investigation on the mass ratios between the contact binaries and their very wide companions may be able to constrain this scenario.

If the Kozai-Lidov mechanism is a dominant formation channel for contact binaries, we would expect them to have companions with smaller separations to trigger the Kozai-Lidov oscillations. Then following the same argument as Section~\ref{sec:result-expected}, we would expect a lower very wide companion fraction for contact binaries because of their enhanced close companion fraction. Instead, our result shows that contact binaries have an enhanced very wide companion fraction, suggesting that either the Kozai-Lidov mechanism is not a dominant formation channel for contact binaries, or there is another mechanism producing a significant number of very wide companions around contact binaries that compensate for the effect of the enhanced close companion fraction.  

To sum up, these very wide companions play a minor role in the orbital migration of the inner systems unless they have undergone a significant outward orbital migration. Therefore, the very wide companions are more likely to be indicative of their formation environment.

\subsection{Formation environment of very wide companions, hot jupiter hosts, and contact binaries}

Because the binding energy of very wide binaries is small, they are sensitive to the environment of their birth place. Several mechanisms have been proposed for the formation of wide binaries. Turbulent core fragmentation may be able to form binaries with separations from a few hundred to a few thousand AU \citep{Offner2010,Lee2017}. Binaries with separations of $10^3$-$10^5$\,AU can form through the dissolution of star clusters \citep{Kouwenhoven2010,Moeckel2011}, the disintegration of unstable compact triples \citep{Reipurth2012}, and pairing of adjacent pre-stellar cores \citep{Tokovinin2017}. In terms of timescales, turbulent core fragmentation and the paring of adjacent cores take place at an age of $\lesssim1$\,Myr during the pre-stellar phase. It takes a longer time (from 10 to a few hundred Myr) for a cluster to dissolve and for a compact triple to unfold.

Wide binaries may be disrupted over time through the gravitational interaction with closely passing stars, molecular clouds, invisible objects, and the Galactic tides \citep{Heggie1975, Bahcall1985,Weinberg1987, Chaname2004, Jiang2010a}. Such disruption takes place on timescales of several Gyr, and most binaries with separations $<10^4$\,AU are not disrupted within the age of the Milky Way \citep{Weinberg1987, Andrews2012}. Even if the wide binaries are disrupted, two stars can still stay in an unbound comoving pair at separations of $\sim100$\,pc for several Gyr because of the small relative velocity \citep{Jiang2010a,Oh2017}. Therefore, most of the very wide companions investigated here are stable over the age of the Milky Way and the disruption events play a relatively minor role.

%Once the wide binaries with separations $<10^4$\,AU ($\sim0.1$\,pc) are formed, most of them are stable over the age of the Universe, although systems with wider separations are more likely to be disrupted due to the perturbations like passing stars and the Galactic tides. 
%Binaries with separations $\lesssim 1$\,pc are mostly stable over time \citep{Weinberg1987}

%Wide binary fraction may also be indicative of their formation environment. For wide binaries forming from the dissolution of star clusters, the wide binary fraction may depend on their cluster mass and size \citep{Kouwenhoven2010}. Dense star formation regions may be difficult for wide binaries to survive \citep{Bate2014}.

The enhanced very wide companion fraction around contact binaries suggests that (proto-)contact binaries are more likely to form in the environments that produce wide systems. For the scenario where wide companions are formed from the dissolution of star clusters, it means that the formation of contact binaries is sensitive to the cluster properties \citep{Kouwenhoven2010}. If wide companions are formed from the disintegration of compact triples, then it implies that  (proto-)contact binaries may be the product of such formation. If the very wide companions are formed from the enhanced turbulent core fragmentation due to certain environmental properties (which may also tend to produce compact multiples), then it suggests contact binaries are also more likely to form in such environment. 

In contrast, the weak or no enhancement of very wide companion fraction around hot jupiter hosts suggests that hot jupiter formation has different dependence on the formation environment as the contact binaries. \cite{Moe2019b} use the different wide companion enhancements between hot jupiter hosts and close binaries to support the idea that hot jupiters are formed from core accretion and (sub-)stellar objects are formed from gravitational instability, coinciding with the mass gap of 4-9\,$\rm M_J$ found by \cite{Schlaufman2018}. The very wide companions investigated here are more sensitive to the birth environment, and the weak or no enhancement of very wide companion fraction around hot jupiter hosts indicates that hot jupiter host formation may be insensitive to larger-scale properties of the birth environment, including the cluster properties and the efficiency of turbulent core fragmentation.

\subsection{The frequency of dual hot jupiter hosts and double contact binaries}

%The occurrences of dual hot jupiter hosts and double contact binaries among the very wide comoving pairs seem to be higher than the occurrence rates of hot jupiters and contact binaries in the field. 

The probability of finding a hot jupiter host (contact binary) in the companion of a hot jupiter host (contact binary) seems to be higher than the occurrence rate of hot jupiters (contact binaries) in the field. Although the sample is small, we find one dual hot jupiter host (two hot jupiter hosts) among 22 hot jupiter hosts that have very wide companions. This $\sim9$\% of hot jupiter occurrence rate (we double count the dual hot jupiter host because that preserves correct statistical properties for inference) in the comoving companions of hot jupiter hosts is much higher than the $0.5-1$\% occurrence rate in the field \citep{Mayor2011,Wright2012,Fressin2013,Santerne2016,Zhou2019}. Similarly, we find 2 double contact binaries out of 188 contact binaries that have very wide companions. This $\sim2$\% occurrence in the very wide companions is also significantly higher than the $0.1$\% occurrence in the field using the same selection method \citep{Hwang2020b}. Our results hint that the occurrence rate of hot jupiters and contact binaries in the comoving companions may be about one order-of-magnitude higher than that in the field. In other words, the occurrence rates of dual systems are higher than the expected values from random pairing of field stars. However, the current sample size of dual systems is still very small so future larger samples are needed to further confirm these results.
%Although the uncertainties are large due to the small sample size of dual systems, the chance of finding a hot jupiter (contact binary) in the companions of a hot jupiter hosts (contact binary) is about one order-of-magnitude higher than that in the field. 

\cite{Tokovinin2014b} also finds an enhanced occurrence rate of 2+2 systems (quadruples consisting of two close stellar binaries) and suggests that these systems were formed by some special process. The disintegration of dynamically unstable compact multiples \citep{Reipurth2012} may also help the formation of double contact binaries, but not for dual hot jupiter hosts. Here we propose another scenario where the enhanced occurrence rate of dual systems is due to the co-chemical (components have similar metallicities) and the co-eval (components have similar ages) nature of the components of wide binaries. \cite{Andrews2018} show that the components of wide binaries with separations $<4\times10^4$\,AU have similar metallicities and elemental abundances within measurement uncertainties (see \citealt{Kamdar2019} for larger separations).  Therefore, if we find a wide companion around a hot jupiter host, then because hot jupiter hosts tend to have higher metallicities \citep{Gonzalez1997,Santos2004,Fischer2005} and the components of wide binaries have similar metallicities \citep{Andrews2018}, we would expect that the wide companion of a hot jupiter host also has a higher metallicity and therefore a higher chance of hosting a hot jupiter, resulting in an enhanced occurrence rate of dual hot jupiter hosts. The close binary fraction is dependent on the metallicity \citep{Moe2019}, and the contact binary fraction is also a function of the stellar age due to their orbital migration and merger \citep{Hwang2020b}, and therefore such metallicity and age dependence of contact binaries can also result in the higher occurrence rate of dual contact binaries.

\section{Conclusions}
\label{sec:conclusion}
In this paper we investigate the very wide comoving fractions with separations of $10^3-10^4$\,AU around hot jupiter hosts and main-sequence contact binaries using Gaia DR2. We further compute the enhancement of very wide companion fractions by comparing with their matched field star samples. We present the following findings:
\begin{enumerate}
	\item $11.9\pm 2.5$\% of hot jupiter hosts and $14.1\pm 1.0$\% of contact binaries have companions at separations of $10^3-10^4$\,AU. Compared to the matched field star samples, the very wide companion fractions are enhanced by a factor of $1.9\pm0.5$ and $3.1\pm0.5$ for hot jupiter hosts and contact binaries, respectively (Table~\ref{tab:result}). 
	\item The measured fraction of very wide companions for hot jupiter hosts is consistent, within $\sim1 \sigma$, with that for matched field stars once we take into account the observational bias in the comparison sample introduced by the lack of close companions to hot jupiter hosts. In contrast, the strong enhancement of very wide companions around contact binaries is highly statistically significant, and there must be a physical mechanism connecting the inner short-period binary with its very wide companion.
	\item We argue that the very wide companions are indicative of the formation environments. The enhanced very wide companion fraction around contact binaries suggests that contact binary formation is sensitive to their formation environment, e.g. the star cluster properties, the efficiency of fragmentation, and/or compact multiples. The weak or no enhancement of very wide companion fraction around hot jupiters implies that the formation of hot jupiters is more tied to their host-star properties instead of large-scale formation environments.
	\item The probability of finding a hot jupiter host (contact binary) in the companion of a hot jupiter host (contact binary) seems to be about an order of magnitude larger than the occurrence rate of hot jupiters (contact binaries) in the field, which may be due to the underlying metallicity and age dependence of hot jupiters and contact binaries. Larger samples are needed to better quantify such occurrence rates. 
\end{enumerate}

%The separations of very wide companions are too wide to induce orbital migration for proto-hot jupiter systems and proto-contact binaries through the Kozai-Lidov mechanism, 

%We find that hot jupiter hosts have a weak or no enhancement of very wide companions, while contact binaries do have a significantly enhanced very wide companion fraction. Quantitatively, $11.9\pm 2.5$\% of hot jupiter hosts have companions in separations of $10^3-10^4$\,AU, and $14.1\pm 1.0$\% of contact binaries have such wide companions. The slightly higher very wide companion fraction of hot jupiter hosts compared to that of the field stars is consistent (at $\sim1\sigma$) with the lack of the close companions around hot jupiter hosts. Our result supports that hot jupiter hosts and the contact binaries undergo different formation processes, and the very wide companions do not affect the formation of hot jupiters. 

The authors are grateful to the  anonymous referee for the constructive report which helped improve the paper. HCH and NLZ thank Scott Tremaine who suggested this work and the Institute for Advanced Study for hospitality. HCH was supported by Space@Hopkins. JH is supported by the Maryland Space Grant.

%\cite{Moe2019b} also consider the separation regime of 50-2000\,AU and show that there is no significant enhancement of having a wide companions compared to the field stars. The discrepancy between our result and \cite{Moe2019b} might be due the different separations being considered.

\bibliography{paper-Hotjupiter}{}
\bibliographystyle{aasjournal}

%\appendix
%Fig.~\ref{fig:contamination-test} presents the test of the contamination on the selection of very wide comoving companions for contact binaries. Similar tests have been applied to hot jupiter hosts and field star samples. In this test, we change the sign of the galactic latitudes of contact binaries, and we also change the sign of the proper motions in the direction of galactic latitudes (after removing the solar motion and the Galactic differential rotation). Therefore, all nearby stars in this test are physically unrelated, chance projection stars. Then we proceed with our comoving companion search, and  Fig.~\ref{fig:contamination-test} shows that no chance project star falls into our comoving candidate selection (the solid demarcation line). 

%\begin{figure}
%	\centering
%	\includegraphics[width=.5\linewidth]{20200224_vel_sep_EB_inverted_b.png}
%	\caption{The test of the contamination by changing the sign of the Galactic latitudes of contact binaries. No chance projection star falls into our comoving companion selection.}
%	\label{fig:contamination-test}
%\end{figure}

\end{document}